\documentclass{sig-alternate} 

\usepackage[T1]{fontenc} 
\usepackage[utf8]{inputenc} 

\usepackage[]{url}
\usepackage[]{mdframed}
\usepackage[]{tabularx}
\usepackage[]{graphicx}
\usepackage[table]{xcolor}
\usepackage{listings} 
\usepackage[linesnumbered]{algorithm2e}
\usepackage{framed}
\usepackage{paralist}
\usepackage{numprint}
\usepackage{color}
\usepackage{array}
\usepackage[justification=centering]{caption}
\usepackage[inline]{enumitem}
\usepackage{comment}

\newtheorem{definition}{Definition}
\newcommand{\nbprograms}{\numprint{6}\xspace}
\newcommand{\nbsosies}{\numprint{24583}\xspace}

\title{Automatic Software Diversity in the Light of Test Suites}

\numberofauthors{1}
\author{          
Benoit Baudry$^o$, Simon Allier$^o$, Marcelino Rodriguez-Cancio$^{\dag\dag,o}$ and Martin Monperrus$^{\dag,o}$\\
$^o$ Inria, France\\ 
$^{\dag\dag}$ University of Rennes 1, France
$^\dag$ University of Lille, France\\
contact: benoit.baudry@inria.fr}     

\begin{document}
\maketitle

\begin{abstract}
A few works address the challenge of automating software diversification, and they all share one core idea: using automated test suites to drive  diversification.
However, there is is lack of solid understanding of how test suites, programs and transformations interact one with another in this process. 
We explore this intricate interplay in the context of a specific diversification technique called ``sosiefication''.

Sosiefication generates sosie programs, i.e., variants of a program in which some statements are deleted, added or replaced but still pass the test suite of the original program. 
Our investigation of the influence of test suites on  sosiefication  exploits the following observation:  test suites cover the different regions of  programs in very unequal ways.
Hence, we hypothesize that sosie synthesis has different performances on a statement that is covered by one hundred test case and on a statement that is covered by a single test case.
We synthesize \nbsosies sosies on \nbprograms popular open-source Java programs.
Our results show that there are two dimensions for diversification.
The first one lies in the specification:  the more test cases cover a statement, the more difficult it is to synthesize sosies. Yet, to our surprise, we are also able to synthesize sosies on highly tested statements (up to 600 test cases), which indicates an intrinsic property of the programs we study.
The second dimension is in the code: we manually explore dozens of sosies and characterize new types of forgiving code regions that are prone to diversification.
\end{abstract}

\section{Introduction}
\label{sec:intro}

Software diversity, i.e., the availability of multiple variants of a program that provide the same functionality with different implementations, is of great interest for software engineering. 
The early exploitation of such diversity was for fault-tolerance in critical software systems \cite{avizienis84,randell75}. 
More recently, the existence of multiple, diverse versions of the same function has been exploited for survivable architectures \cite{knight2002}, cross-checking oracle \cite{Carzaniga14}, self-adaptation \cite{Ghezzi13},  intrusion detection  \cite{wang03} and multi-level diversification  \cite{allier14}.

As opposed to the exploitation of manually created software diversity as in N-version programming \cite{avizienis84}, there is a research area on automatic software diversity, ignited by the seminal works of Cohen \cite{cohen93}  and Forrest \cite{forrest97}.  
Automatic diversification has been widely explored at machine-code level for security purposes \cite{baudry2015}, but only a few works tackle this challenge in application-level source code \cite{rinard2005,Schulte13,langdon2013,baudry14}. 
They all share the same core idea: using automated test suites to drive diversification.
In short, the process consists of transforming the original program to get a variant and of running the test suite to assess the validity of the variant.
\emph{However, there is is lack of solid understanding of how test suites, programs and transformations interact one with another in this process. There lies our contribution.}

In this work, we consider a specific diversification technique  called ``sosiefication'' \cite{baudry14}.
Sosiefication creates \emph{sosie programs} that are variants of a program in which some statements are deleted, added or replaced but still pass the test suite of the original program. 
Our intuition is that the test suite of a program, the basis for all recent works on automatic diversification, covers the different regions of the program in very unequal ways, and that it has an impact on sosiefication.
We hypothesize that synthesizing a sosie on a statement that is covered by one hundred test case is different from synthesizing a sosie on a statement that is covered by a single test case.
The difference lies in the ease of synthesis and in the quality of the resulting sosie.
This is what we explore in this paper.

Technically, we synthesize \nbsosies sosies on \nbprograms popular open-source Java programs that are available with very solid JUnit test suites.
For each of them, we compute all ``execution signature'' per statement, a short expression that refers to the number of test cases that execute a given code region. 
We consider this metric as a proxy to the ``amount of specification'' -- so to speak -- of this region
We show that this metric greatly varies for the statements inside a program.
We use this metric as guiding light for our investigation of the mechanisms that underlie  sosiefication.

We show that there is a relation between execution signatures and the efficiency of the sosiefication process:
the more a statement is tested, the more difficult it is synthesize sosies.
However, to our surprise, we are still able to synthesize sosies on highly tested statements (up to 600 test cases).
To us, this indicates an intrinsic property of the software subjects under study. 

In addition to a quantitative analysis on the sosiefication process, we perform a qualitative investigation of sosiefication via manual assessment.
We propose a first categorization of sosies, where each category relates to a specific kind of code region (e.g. optimization code). 
This extends the body of knowledge about forgiving code regions \cite{rinard2012}.
In particular, we find regions characterized by ``plastic specifications'', i.e. regions  which are governed by a very open yet strong contract. For instance, the only correctness contract of a hashing function is to be deterministic. On the one hand this is a strong contract. On the other hand, this is very open: many variants of an hashing function are valid, and consequently, many modifications in the code result in valid hashing functions.

We believe that our findings based on a specific diversification technique -- sosiefication -- can be exploited for other  diversification approaches.
We provide novel insights about two dimensions of diversification.
First, we shine a spotlight on  the existence of plastic parts in program specifications.
The literature has already identified some, e.g., video compression and in this paper we reveal a new one based on hashing function. 
But we are convinced that there are many other such plastic specifications. \emph{Future research has to build a comprehensive catalog of plastic behavior.}

The second dimension is in the code. The forgiving regions parts  of the code  are those that can be easily modified while maintaining acceptable behavior.
Often, the implementation of plastic specifications are forgiving (such as the implementation of a video codec). 
However, this is not a bijection. In our manual analysis, we have encountered forgiving statements in zone that are every conventionally binary in their specification. \emph{There is a need for research on the intersection of plastic behavior and forgiving regions.}

To sum up, the contributions of the paper are:
\begin{itemize}[leftmargin=.4cm,itemsep=-.1cm]
  \item an empirical analysis of the interplay between  programs and their test suites that demonstrates the wide variety of execution signatures
  \item quantitative evidence of the relation between the uneven coverage of statements and the opportunities for automatic program transformations
  \item a deeper understanding of forgiving code regions that can be exploited for sosiefication as well as for other forms of automatic diversification (as targets for automatic transformation).
\end{itemize}

The paper is organized as follows.
Section \ref{sec:preliminary} presents a preliminary  analysis that demonstrates uneven coverage of different regions of a program by its test suite. 
Section \ref{sec:methodo} recalls the essentials about sosie synthesis, as well as our experimental protocol. 
Section \ref{sec:results} presents and discusses our main findings about the interplay between a test suite, a program and the opportunities for sosie synthesis.
Section \ref{sec:rw} outlines the related work and section \ref{sec:conclusion} concludes.

\section{A Preliminary Study about statement execution signatures}
\label{sec:preliminary}

In this paper, we are interested in how test suites, programs and transformations interact.
In this section we explore the relation between the first two: test suite and programs.
We  perform a preliminary experiment about the interplay between the test suite of a program and its statements.
We consider projects written in Java and coming with a JUnit test suite.
In this, test code is clearly separated from application code and each test case includes one or more method calls, and one or more assertions that express the expected properties about the program's behavior.

\begin{figure*}
  \centering
  \includegraphics[width=0.92\textwidth]{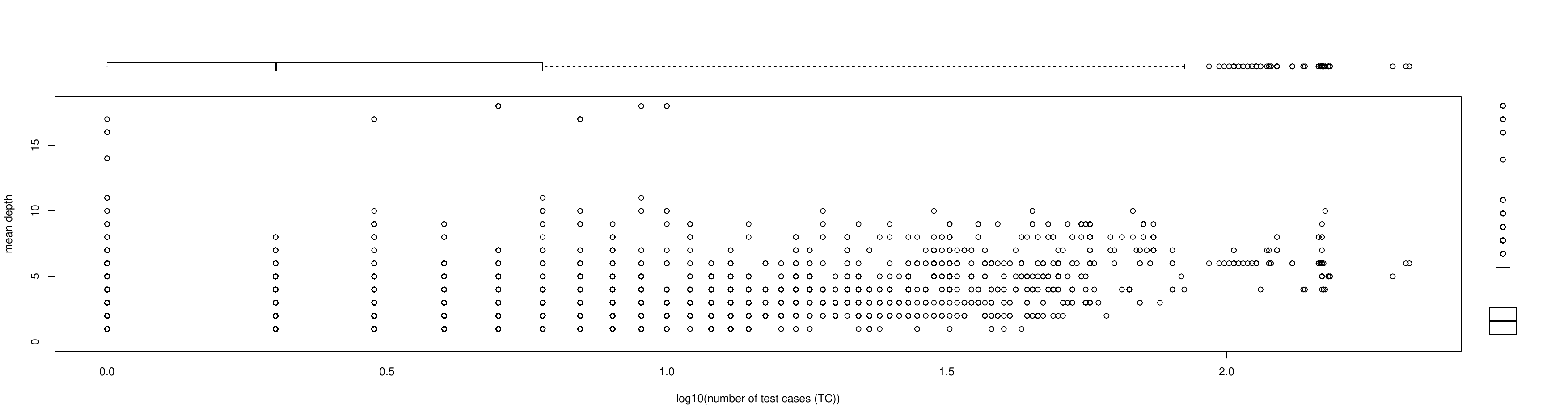}
  \caption{Interplay between program statements and test suites in Apache Commons Lang. A point is a statement covered by $n$ test cases, where $n$ is the X-axis. The Y-axis is the  mean depth  of the statement  when running the whole test suite.}
  \label{fig:cov-distribution}
\end{figure*}

\subsection{Collecting Statement Execution Signatures}

We have developed a tool, called SESig, which collects fine-grained metrics about how the statements in a Java program are covered by a test suite. 
It collects the following metrics about each statement:
\begin{enumerate*}
  \item the number of test cases in the test suite that cover the statement s;
  \item The execution depth of s. We associate a vector $Depth_s$ to each statement, such that, given the set $\{t_1, ... t_n\}$  of test cases that cover s $Depth_s=[depth(s,t_i)]_{i \in [O..n]}$. 
  $depth(s,t_i)$ is the depth of s in the call stack when running $t_i$.\footnote{ \scriptsize
  When counting the depth in the call stack, we ignore calls to external libraries.}.
\end{enumerate*} 

For example, let us consider the method \texttt{append} from commons.lang 3.3.2  (Listing \ref{lst:lang-example}). 
SESig collects the following information. 
The method is executed by 28 different test cases and all statements of the method but one (line \ref{notcovered}) are covered. 
Most statements are executed by one test case only, except the two statements in lines \ref{multitest1} and \ref{multitest2} that are executed by 24 and 25 different test cases respectively. 
We also observe that most statements are executed at depth 1 except ones in lines \ref{deep1} and \ref{deep2} that are executed only at depth 6.
Listing \ref{lst:stack} shows the stack trace when stopping on this statement: we clearly see that they are not directly  exercised by a test case.
Statements in lines \ref{multitest1} and \ref{multitest2} appear at different depths, indicating that the different test cases that cover them trigger these behaviors in different contexts.

\begin{lstlisting}[caption={The \texttt{append} method from \texttt{FieldUtils} in commons.lang}, label=lst:lang-example, language=java, numbers=left]
public EqualsBuilder append(final boolean[] lhs, final boolean[] rhs) {
  if (isEquals == false) {
    return this;}           //(0,[]) %* \label{notcovered} *)
  if (lhs == rhs) {
    return this;}           //(1,[1])
  if (lhs == null || rhs == null) {
    this.setEquals(false);  //(1,[1])
    return this;}           //(1,[1])
  if (lhs.length != rhs.length) {
    this.setEquals(false);  //(1,[6]) %* \label{deep1} *)
    return this;}           //(1,[6]) %* \label{deep2} *)
  for (int i = 0; i < lhs.length && isEquals; ++i) {
    append(lhs[i], rhs[i]); //(24,[1,2,5,6])  %* \label{multitest1} *)
  }
  return this;}             //(25,[1,2,3,5,6]) %* \label{multitest2} *)
\end{lstlisting}

\begin{lstlisting}[caption={Stack trace when stopping at line \ref{deep1} of Listing \ref{lst:lang-example}}, label=lst:stack, language=java, numbers=none]
EqualsBuilder.append :899
EqualsBuilder.append :487
EqualsBuilder.reflectionAppend :411
EqualsBuilder.reflectionEquals :360
EqualsBuilder.reflectionEquals :295)
DiffBuilder.<init> :111
DiffBuilderTest.testBooleanArray :110
\end{lstlisting}

SESig adds probes in the test suite and the program, at the following locations: entrance and exit of a test case, entrance and exit and methods in the program, bifurcation of branches inside a method, each statement in the program (this latter probe collects the depth of the statement in the call stack and id of the test case that is currently running). 
The tool is publicly available as open source \footnote{ \scriptsize \url{github.com/DIVERSIFY-project/sosies-generator/tree/icse15}}.

\subsection{Empirical Observations}

We now explore the test suite execution at the level of an entire project.
Figure \ref{fig:cov-distribution} displays the signatures of all statements in Apache Commons Lang that are covered by one test case at least. Each point is a statement and its position indicates the number of test cases that cover it (X-axis) and its median  depth in the call stack  when running the test cases (Y-axis). 

The x-axis captures the disparity in terms of coverage, summarized in a boxplot at the top of the figure: some statements are covered by no more than one test case (4243 statements), while some others are covered by hundreds of test cases (77 statements are covered by more than 100 test cases). 
Yet, test coverage is very skewed towards low values: 25\% of the statements are covered by a single test case and 50\% are covered by one or two test cases.

The y-axis captures the disparity in the relative position of a statement in the execution flow of a test suite: a majority of statements are executed close to the test case (at a depth lower than 5), while some others appear much deeper and are most probably tested only as a side-effect of testing other methods. 
For example, statement at line \ref{deep1} of Listing \ref{lst:lang-example} appears at a depth of 6 calls in the stack and is not the main testing target of the single test case that covers it. 
What clearly appears here is that a vast majority of statements appear quite close to the test cases that cover them (75\% of statements have a median depth below 2.5). 

We manually looked at the extreme cases. 
The statements that appear very deep in the stack (more than 13, on the top part of figure \ref{fig:cov-distribution}) are statements in recursive methods. 
These have a high median depth value and also very large variance in their depth value: all of them happen to be actually tested at depth 1 as well as at depth above 30. 
Looking at the statements that are covered by many test cases (on the right of the plot), we remark that they are also always at a median depth greater than 1. 
These statements are mostly in utility methods that are used by many other methods, hence all of them are both directly tested and indirectly tested through the test case of client code (e.g., the right-most statements are all in the \texttt{ToStringStyle} class). 


We performed the analysis for other programs that will be used later  in this paper and presented in Table \ref{tab:subjects}). All plots are available online\footnote{\scriptsize \url{github.com/DIVERSIFY-project/sosie-results}}. 
The maximum values for the number of covering test cases  and median depth vary from one project to the other: the most covered statement of commons.codec is covered by 105 test cases, while the maximum of commons.collection is 1780 test cases that cover a statement; the median depth varies from 1 to 8 in commons.io and from 1 to 1863 in GSon. 
Yet, some major trends are observed in all projects: 
\begin{inparaenum}[(i)]
  \item statements are always very unequally covered by the test suite;
  \item 50\% of the statements are covered by a small number of test cases: this number varies between 2 (as in the case of lang) and 11 (in GSon);
  \item the statements that appear very deep in the execution stack are always in recursive methods (the most extreme cases were observed in GSon, where some statements appeared as deep as 3692); and
  \item the statements that are covered by a large number of test cases occur at a mean depth greater than 2 because they are in utility methods or in private methods, hence mostly executed  by methods in the program rather than directly by the test cases.
\end{inparaenum}

\begin{framed}
To sum up, this preliminary study suggests that the statements have very different execution signatures.
Our intuition is that we can leverage these large variations among signatures to characterize the interplay between test suites and program statements for software diversification. 
\end{framed}

\section{Analysis of a Diversification Technique}
\label{sec:methodo}

We have observed that the interplay between a test suite and the the statements of the program under test produces very different statement signatures.
Our goal is now to relate these statement signatures to a particular diversification technique: sosiefication.

\subsection{Sosie synthesis}
\label{sec:background}

Sosiefication is the process of synthesizing sosies.
We have introduced it in our previous work on software diversity \cite{baudry14}. The word sosie is a French word that literally means ``look alike''.

\begin{definition}\label{def:sosie} \textbf{Sosie} (noun).
  Given a program $P$, a test suite $TS$ for $P$ and a program transformation $T$, a variant $P'$=$T(P)$ is a sosie of $P$ if the two following conditions hold
1) the part of $P$ that is modified by $T$ is covered by one test case at least;
2) all test cases in $TS$ pass on $P'$.
\end{definition}

Given an initial program, we synthesize sosies with source code transformations that  modify the abstract syntax tree (AST).
We consider three types of transformation that manipulate statement nodes of the AST:
1) remove a node in the AST (Delete);
2) adds a node just after another one (Add);
3) replaces a node by another one from the same AST (Replace). 
We call the \textbf{transplantation point} the statement on which we perform a transformation.
For \texttt{add} and \texttt{replace}, we also refer to the \textbf{transplant statement} that is copied and inserted. 
The transplantation and transplant points are in the same AST (we do not synthesize new code, nor take code from other programs). 

Sosiefication consists in randomly picking an AST statement node and try to apply the three transformations.
Yet,  for \texttt{replace} and \texttt{add}, we introduce some constraints.
First, a statement cannot be replaced by itself; AST nodes of type \emph{case},  \emph{variable declaration}, \emph{return} and \emph{throw} are only replaced by statements of the same type; the type of the value returned by a \emph{return} statement must be the same for the original and new statement.
Second, we consider transplant statements that manipulate variables of the same type as the transplantation point, and we rename the variables of the transplant with names of variables of the corresponding type, which are in the namespace of the transplantation point. 
We call this \emph{Steroid} transformations \cite{baudry14}.

Since the sosiefication process consists in applying a transformation on a program and then running the test suite to select sosies, it can look similar to mutation testing. 
Sosies might even be thought of as equivalent mutants. 
Yet, both approaches are conceptually different: program transformations for mutation testing are designed according to fault models, while the sosiefication transformations are designed to explore the neighbourhood of similar programs; mutation testing aims at assessing the ability of a test suite at detecting the injected bugs, while sosieficiation aims at synthesizing variants of a program that exhibit a form of diversity. 
Also, we have shown that, by opposition to equivalent mutants, sosies can behave differently from the original and produce different results under certain conditions  \cite{BaudryARM15} (and we illustrate more examples in  section \ref{sec:taxonomy}).

\subsection{Metrics}
\label{sec:metrics}

We now present a metric that characterize the sosiefication process, as well as  the  features that characterize a transplantation point in which sosiefication can be applied.

\begin{definition}\label{def:SR}      \textbf{Sosiefication Rate (SR) }                                                 
is the ratio between the number of sosies (variants that pass the test suite), and the total number of transformations done, one transformation being a trial to produce a program variant:\\ $\# Sosies/\# Trials$.
\end{definition}

Sosiefication is an expensive process, which uses a lot of computation power. 
From an engineering perspective, it is good to generate as many sosies as possible in any given amount of time.
To this extent, it is better to maximize the sosiefication rate.

Our goal is to explore the relations between transplantation points and  the sosiefication rate. 
For instance, we are especially interested in the transplantation point features that maximize the sosiefication rate. 
We focus on the following features to characterize transplantation points.  

\begin{definition}\label{def:features}   
  \textbf{Transplantation point features}:    
Let us call $\tau$ the transplantation point yielding the sosie. We focus on the following features:
\begin{inparaenum}[\itshape 1\upshape)]
  \item$TC_{\tau}$ is the number of test cases that execute $\tau$.
  \item$Transfo_{\tau}$ is a categorical feature that characterizes the type of transformation that we performed on $\tau$: add, delete or replace. This can be further refined by considering the type of AST node where the transformation occurs.
\end{inparaenum}
\end{definition}

The collection  of all those features is implemented in a tool that is publicaly available \footnote{\scriptsize \url{https://github.com/DIVERSIFY-project/sosies-generator/tree/icse15}}.

\subsection{Experimental Protocol}
\label{sec:protocol}

In this paper, we perform the following experiment.
For a set of programs considered as a dataset (presented in table \ref{tab:subjects}), we synthesize a set of sosies. For this, we use the ``Steroid'' strategy as described in section \ref{sec:background}. 
This process is budget based: we try neither to exhaustively visit the search space nor to have a fixed-size sample. 
Since sosiefication is an expensive process, our computation platform is Grid5000, a scientific platform for parallel, large-scale computation \cite{bolze2006grid}.
We submit one batch for each program, it is run as long as resources (CPU and memory) are available on the grid.
Then, for each sosie, we extract or compute the metrics described in previous section.
We  also manually analyze dozens of sosies in order to build a taxonomy of sosies. 

\begin{table}[ht]
\centering
\caption{Descriptive statistics about our subject programs}
\begin{tabularx}{\columnwidth}{lXXXX}
\hline
& \#classes & \#stmt & \#TC & cov.   \\
\hline
commons-lang 3.3.2 & 132 & 8442 & 2352 & 94\% \\
commons-collections 4.0  & 286 & 6780 & 13677 & 84\% \\
commons-codec 1.10  & 60 &2695 & 662 &  96\% \\
commons-io 2.4 & 103& 2573& 962&  87\% \\
Gson 2.3.2& 66& 2377& 951&  79\% \\
jgit 3.7.0 & 666 &22333 &2758 & 70\%\\
\hline
\end{tabularx}
\label{tab:subjects}
\end{table}

We consider the \nbprograms programs presented in table \ref{tab:subjects}. 
All  programs are popular Java libraries developed by the Apache foundation, Google or Eclipse.
The  second column gives the number of classes,
the third column the number of statements.
Column 4 provides the number of test cases executions when running the test suite and column 5 gives the statement coverage rate.

The programs range between 60 and 666 classes. 
All of them are tested with very large test suites that include hundreds of test cases that execute the program in many different situations. 
One can notice the extremely high number of test cases executed on commons-collection. 
This results from an extensive usage of inheritance in the test suite, hence many test cases are executed multiple times (e.g., test cases that test methods declared in abstract classes). 
The test suites cover most of the program (up to 96\% statement coverage for commons-codec). Jgit is the exception (only 70\% coverage): it includes many classes meant to connect to different remote git servers, which are not covered by the unit test cases (due to the difficulty of stubbing these servers)
This dataset provides a solid basis to investigate the interplay between test suites and sosiefication. 

\subsection{Research Questions}

We contribute to the exploration of two general problems of software diversification: how to effectively synthesize diverse software? what property of software should be searched and exploited for the sake of diversification? 
The following research questions are contributions in this direction.

\subsubsection{RQ1: Is the sosiefication rate SR higher for statements that are less tested (in terms of number of test cases)?}

One criticism often made about techniques that rely on test suites to automatically transform programs \cite{langdon2013,legoues12,baudry2015,rinard2005,schulte2014} is that test suites are not strong enough to ensure the validity of variants.
The intuition behind this criticism is that if a program is badly tested, it is easy to generate variants of the program that still pass the test suite. 
This research question investigates to what extent this is true for sosie synthesis, by comparing the sosiefication rates on poorly test regions with the rates on highly tested regions.

\subsubsection{RQ2: what is the relation between the types of transplantations points and transplants and the test suite execution?}

We would like to understand the interplay between the transformation operators and the test suite. For instance, it may happen that if-conditions are better specified than methods calls.
This has a direct impact on sosiefication, while the sosiefication on if-conditions may yield a higher sosiefication rate, they may also be of worse quality.
There are three dimensions in the qualification of transformations: 
1) how they are applied (addition of new code versus deletion of existing code);
2) where they are applied, i.e. the transplantation points (e.g. ifs versus method calls);
and 
3) for addition and deletion, the type of the transplant.
This research question studies those three dimensions.

\subsubsection{RQ3: What are the different kinds of good sosies that we can generate?}

In our experience, certain sosies are really interesting, and others are ``bad''. The bad ones  are those that are obviously incorrect. These sosies pass the test suite, by construction, but they happen in parts that are loosely specified. 

Meanwhile, our experience also showed that there exists different kinds of good sosies, e.g., sosies that introduce true diversity in the computation and not merely bugs.
This research question relies on the manual analysis of dozens of sosies from all programs of our dataset, to build a taxonomy of good program sosies.

\section{Empirical Results}
\label{sec:results}

We apply our experimental protocol on \nbprograms Java programs.
Table \ref{tab:sosies} gives the key data about the sosies computed with the budget based approach described in \ref{sec:protocol}. 
The second column indicates the number of sosies we generated for each program, 
the third column indicates the global sosiefication rate (SR), i.e., among all variants that we generated how much were actual sosies (the other variants either don't compile or fail for one test case at least), 
the next columns indicate the number of sosies synthesized by adding, deleting or replacing statements, 
the last column indicates the rate of statements for which we generated variants, i.e., the number of statements that served as transplantation points over all statements. This last metric provides an indication of how much we tried to sosiefy in all regions and thus to what extent we can exploit the findings of section \ref{sec:preliminary} to investigate the sosies.  The low rate for jgit is related to large size of our project: since sosiefication has a bounded a resource budget, we cannot cover a large program as much as a small one.

\begin{table}[ht]
\centering
\caption{The Sosie Programs Considered on our Empirical Investigations}
\begin{tabularx}{\columnwidth}{lXXXXXX}
\hline
 & \#sosies  & sosief.  & add &del&rep & expl.   \\
 & & rate (SR) & & & & rate \\
\hline
lang        & 1146  & 9.6\% & 419& 190& 537 & 78\%  \\
collections & 8626  & 10.8\%& 3912& 754& 3960& 83.3\%  \\
codec       & 701   & 10.4\%& 289& 146&266 &  91.9\% \\
io          & 3545  & 13.9\%& 1754& 319&1472 & 92\%  \\
Gson        & 4311  & 14\%  & 2199& 215& 1897& 80.3\%  \\
jgit        & 6262  & 16\%& 1924& 1375& 2963& 57\%  \\
\hline
\end{tabularx}
\label{tab:sosies}
\end{table}

\begin{figure*}[!ht]
  \centering
  \includegraphics[width=0.92\textwidth]{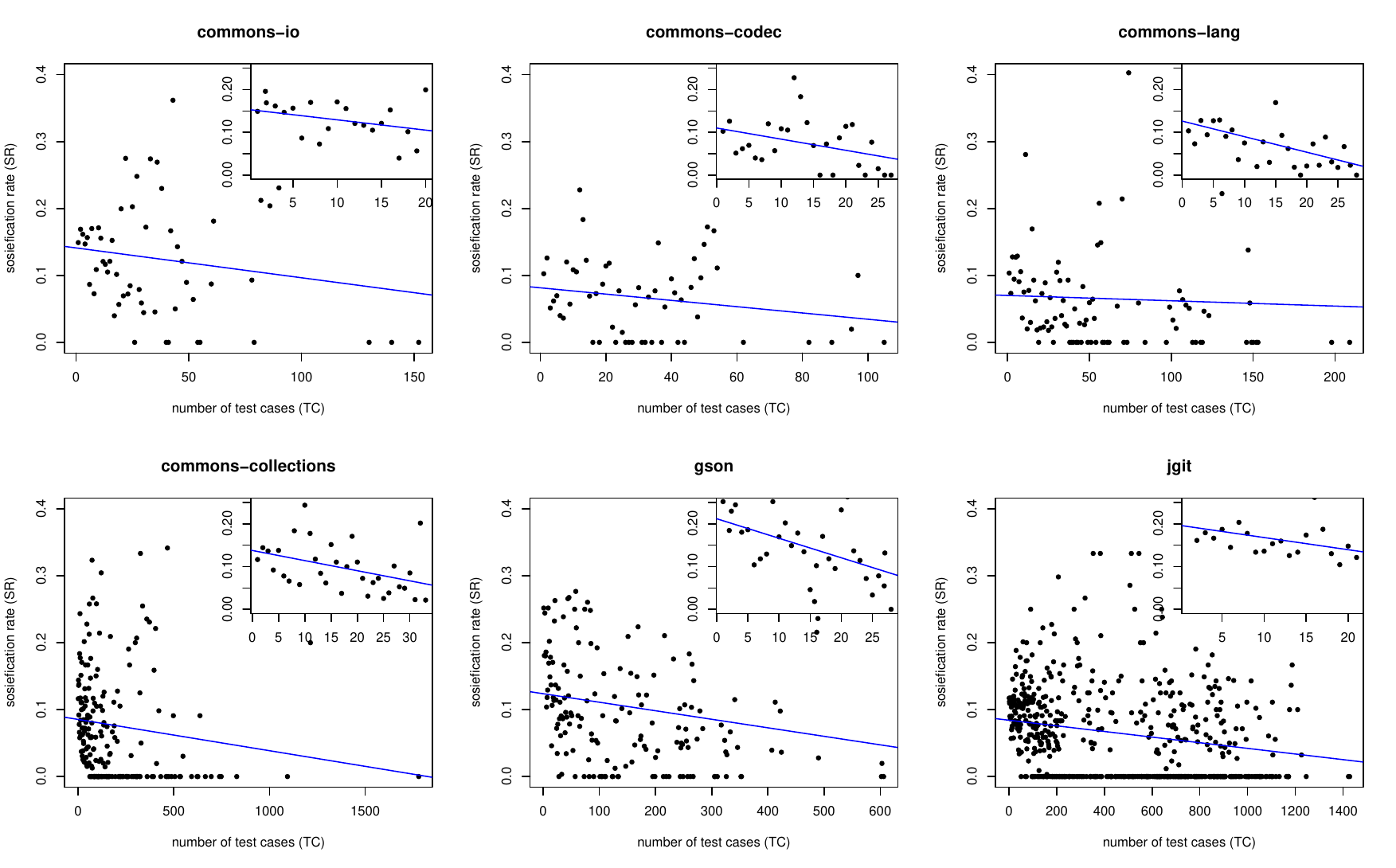}
  \caption{Distribution of sosiefiaction rate w.r.t coverage of the transplantation point: one point on a plot represents the global sosiefication rate at transplantion points covered by a given number of test cases. Each plot includes the best possible linear regression. In the top right corner of each plot, we also include a zoom on the left hand side of the distribution (e.g. from 1 to 20 test cases for Commons IO).
}
  \label{fig:SR-TC}
\end{figure*}

\subsection{RQ1: Relation between Statement Execution Signature and Sosiefication}

We try to apply one or more transformation at each transplantation point, in order to create sosie programs.
Each trial produces a program variant, which either fails at compiling or fails at passing the suite or be a sosie, and we then compute the sosiefication rate (cf. definition \ref{def:SR}) at each transplantation point. 
Since a transplantation point is a statement, we use SESig to retrieve the number of test cases that cover it.


We analyze the cumulative sosiefication rate at transplantion points covered by a given number of test cases. 
Figure \ref{fig:SR-TC} provides this data as scatter plots. 
We have removed the outliers (sosiefication rate that are to high due to degerated cases discussed below).
It contains \nbprograms subfigures, one per project of our dataset.
For instance, the first figure is for Apache Commons io. 
This program includes 845 statements covered by a single test case, 756 of them are optential transplantation points for trying sosie synthesis. The cumulative sosiefication rate for these points is 15\%: we performed a total of 10959 trials on the 756 points and 1644 were actual sosies.

In the top right corner of each plot, we also include a zoom on the left hand side of the distribution (e.g. from 1 to 20 test cases for Commons IO).
The rational for this zoom is that a vast majority of the statements -- hence transplantation points -- are on the left (as shown in section \ref{sec:preliminary}, the distribution of statement coverage is highly skewed towards low values), and this is also where we performed the highest numbers of trials. 

This data can be interpreted as follows. 
First, for all projects, the sosiefication rate tends to decrease with the number of test cases. 
The slope of the decrease varies between $4\times 10^{-5}$ and  $7\times 10^{-3}$ for global plots.
This is a variation of three orders of magnitude.
The slope itself is low because the X-axis is an absolute number of test cases going up to $10^3$ while the Y-axis is by construction between 0 and 1.
The general tendency to decrease can be explained by the fact that more test cases means more testing scenarios and more assertions, which means that this lets less space for unspecified behavior.
Since the sosiefication process heavily explores this space by construction, more test cases directly results in a lower sosiefication rate. 
In other words, the increase in specification quality yield fewer sosies (the buggy program variants being killed).
Interestingly, the decrease in the zooms, i.e. for the poorly tested sosies, is higher with slopes ranging from $2\times 10^{-2}$ to $4\times 10^{-2}$.
This can be interpreted by the accentuation of the ``specification quality improvement'' phenomenon on the left part of the plot: we believe that, in terms of behavioral specification, the ``amount of additional specification'' of a given statement is generally higher between one and two test cases than between 600 and 601. Here, the unconventional expression ``amount of additional specification'' refers to new contracts, new corner cases, etc. 

Second, one sees that the right hand side of the distribution is very irregular. 
For instance, for Apache Commons Collections, we see several spikes  from 0 to 0.4 among points  above 30 test cases .
This can be explained by several factors.
The main one is that sosiefication rate -- a ratio -- has degenerated cases. One degenerated case is the absence of data: for instance, there is no statement that is covered by exactly 131 test cases in program Apache Commons Collections.
Another degenerated case is when there is too few data. For instance, in Gson, there is one single statement which is covered by 372 test cases. By chance, the variant made on this statement is a sosie. Consequently, for n=372 test cases, the sosiefication rate is 100\%. However, the average sosiefication rate for hundreds of test cases is not at all in the 100\%.
This case is clearly an outlier, due to the limited amount of data (as we saw in section \ref{sec:preliminary}, there is only a limited number of statements covered by many test cases).


Beyond this graphical interpretation, we have performed the following statistical test.
For each project, we have manually selected a threshold separating low-tested transplantation points from high-tested transplantation points. This project-dependent threshold\footnote{\scriptsize io: 20, codec: 27, lang: 28, collection: 33, gson: 28, jgit: 21} corresponds to the thumbnails, which show the low-tested points that are below the threshold.
This yields two different sosiefication rates, the sosiefication rate of low-tested transplantation points and the rate for high-tested ones. 
Since a rate is a proportion, we can perform a standard equality-of-proportion test, as implemented by `prop.test' in R. For 4/6 projects, the null hypothesis (``the sosiefication rates are the same'') is rejected with 95\% confidence.
For io and lang, with a respective p-value of 0.4 and 0.08, there is not enough data to reject the null hypothesis.

The third finding is that there are no  project for which the sosiefication rate clearly tends towards zero. In other terms, our data suggests that whatever the amount of specification, our code transformations still produce program variants that are sosies. We explain this by the presence of \emph{software plasticity}, a concept that we introduce in this paper and for which we propose a first characterization. 

We define \emph{software plasticity} as the ability of software modules to have different behaviors while still remaining correct.  \emph{Software plasticity} is very much related to Rinard's work where the transplantation points happen to be in ``forgiving regions'' of code \cite{rinard2012}. 

To some extent, the sosiefication rate when the number of tests is high reflects this amount of software plasticity. It may even be the very first quantitative measure of it. If we put several data points in bins, we smooth the irregularities shown in Figure \ref{fig:SR-TC}. This results in an overall sosiefication rate of 10\% for GSon.
In Rinard's term, the sosiefication rate obtained with our protocol suggests that there exists 10\% of forgiving regions in GSon.

\begin{framed}
Answer to RQ1: Transplantion points covered by few test cases are easier targets for sosiefication. 
However, the sosiefication rate never goes down to zero. To us, it hints to an intrinsic property of the software subjects under study.
We hypothesize that this property is the presence of software plasticity and forgiving regions. 
\end{framed}

\subsection{RQ2: Relation between Transplantation Points, Transplants and Test Suite}

We now look at  whether the different types of program elements (i.e. types of AST nodes) are specified equally. 
Hence, we compute the sosiefication rate per AST node type.

\begin{figure}
  \centering
  \includegraphics[width=\columnwidth]{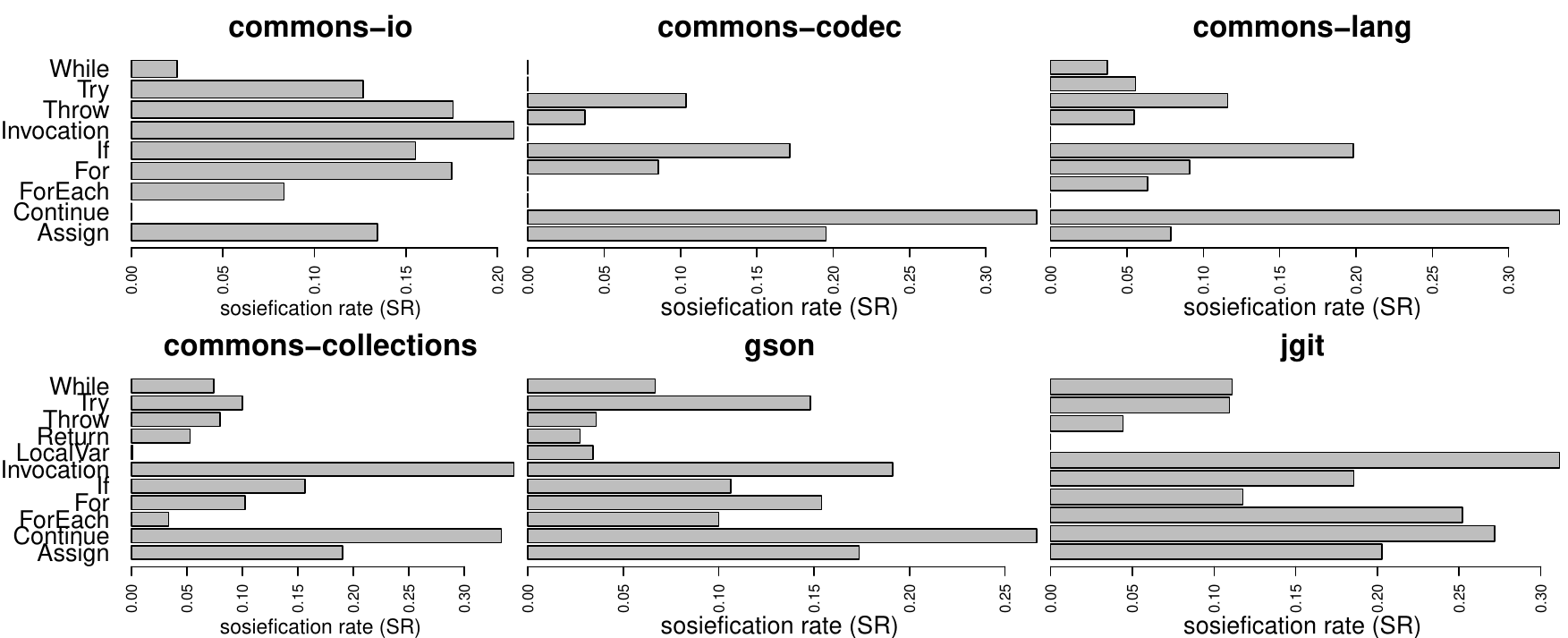}
  \caption{The sosiefication rate for the \texttt{delete} transformations, according to the type of the transplantation point}
  \label{fig:SR-delete}
\end{figure}

We start with the sosiefication operator ``delete'' (based on the number of sosies given in table \ref{tab:sosies}).
Figure \ref{fig:SR-delete} provides the sosiefication rate with \texttt{delete} transformations according to the type of the transplantation point. 
This shows that there is large variation in the sosiefication rate per node type.
For instance, this figure suggests that method invocations are less specified than while-blocks, since the sosiefication rate is higher.

\begin{figure}
  \centering
  \includegraphics[width=\columnwidth]{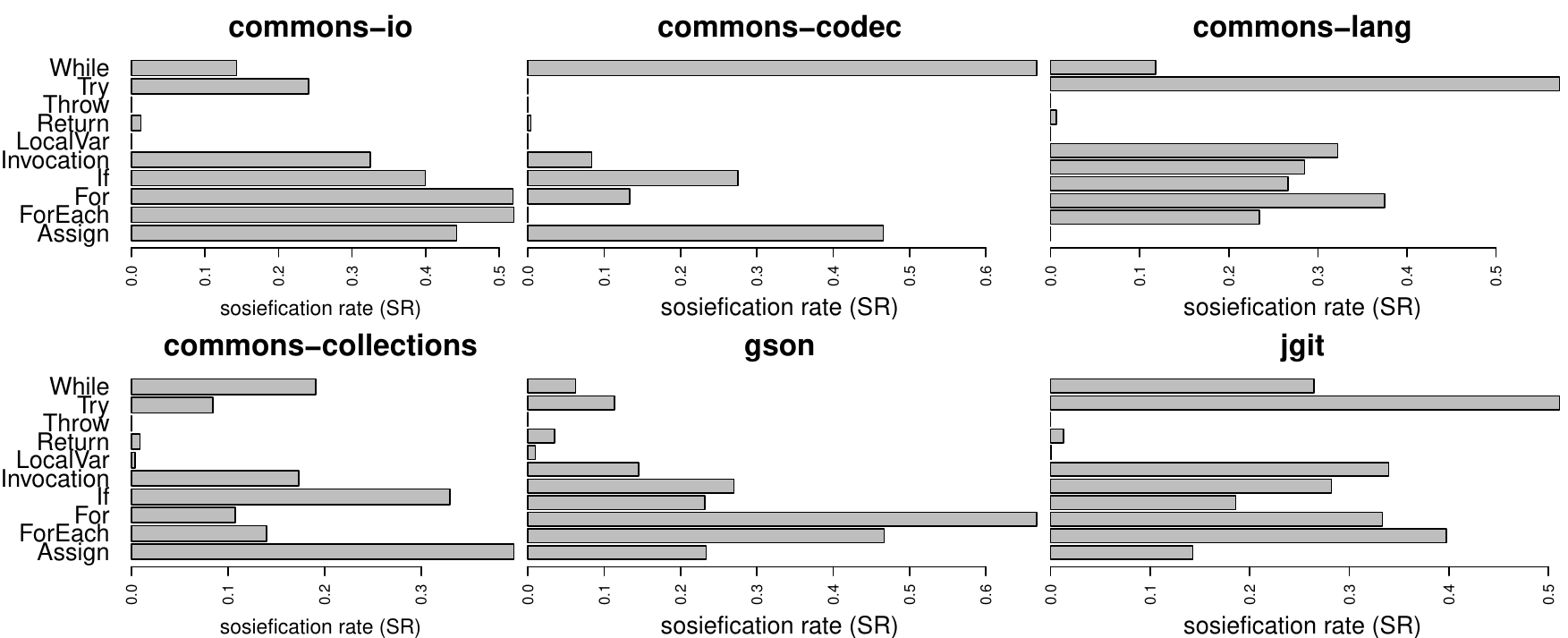}
  \caption{The sosiefication rate for \texttt{add} transformations, according to the type of the transplant.}
  \label{fig:SR-add}
\end{figure}

Considering the sosiefication operator ``add'', figure \ref{fig:SR-add} provides the sosiefication rate according to the type of the added code, i.e. the transplant (and not the transplantation point).
We see that there are also large variations between node types as well as between projects. However, some regularities emerge: for instance, adding a return always yield a low sosiefication rate. Along the line of RQ1, this means that ``return'' nodes are widely specified. This matches the intuition that most assertions in test suites are made on returned values just after the computation.

However, those two figures can be interpreted from a different viewpoint.
Let us consider again Figure \ref{fig:SR-delete} about the sosiefication rate for \texttt{delete} transformations
We can see that the deletion of \texttt{continue} nodes is always the most effective for sosiefication. 
Those nodes are usually used as shortcuts in the computation, hence removing them yields slower yet acceptable program variants; we discuss this in depth in the next section. 
We also observe a good sosiefication rate for deletion of method invocations. 
We explain this effect by the presence of side-effect free methods which can be safely removed (discussed also in the next section) and by the existence of many redundant calls  (discussed in next section). 

The same alternative viewpoint can be taken on code addition.
Looking more closely at figure \ref{fig:SR-add}, we realize that for all projects, the addition of assignment nodes is the most effective. 
This can be explained by the fact that there are many places in the code where the variable declaration and the first value assignment for this variable are separated by a few statements. 
In these situations it is possible to assign any arbitrary value to the variable, which will be cancelled by the subsequent assignment. Yao and colleagues observed a similar phenomenon of specific assignments that ``skeeze out'' a corrupted state \cite{yao14}).
Also, for some project such as commons-io and jgit, the addition of method invocations is also quite effective.
Similarly to deletion, it probably indicates a non-negligible proportion of side-effect free methods in the program.
The addition of conditionals and loops is also effective. It is important to understand that a large number of these additional blocks have conditions such that the execution never enters the body of the block.

Considering \texttt{replace} transformations that combine  deletion and addition, they always  have the lowest sosiefication rate. 
We do not provide any graphical representation of this data, for space constraint reasons. 
Yet, we make the following observations. 
First, picking a transplant and a transplantation point that are method invocations is quite effective. This suggests the presence of alternative yet equivalent calls, that is discussed in the next section and also  by Carzaniga et al. \cite{Carzaniga14}. 
Second, we observe a certain plasticity around \texttt{return} statements: some of them can be replaced by the statement surrounded by a try or a condition. This suggests the existence of similar statements in the neighbourhood of the transplantation point, which perform additional checks.

\begin{framed}
Answer to RQ2: The addition of new statements is always the most effective way to produce new sosies. Deletion is most effective for some AST nodes types such as ``continue'', to some extent, those AST nodes tend to be micro forgiving regions. This new knowledge is actionable for designing the next generation of sosie synthesizer, and maybe leveraged for other diversification techniques.
\end{framed}

\subsection{RQ3: What are the different kinds of good sosies that we can generate?}
\label{sec:taxonomy}

With RQ1, we have seen that the sosiefication rate depends on the test suite execution signature.
Now, we are interested in understanding whether there is a difference in nature between the sosies produced on low-tested transplantation points and those produced on high tested transplantation points.

For each program, we selected sosies among extreme cases: those  synthesized on transplantation points covered by a single test case or synthesized on points covered by the highest number of test cases.
By doing this, we are able to build a taxonomy of sosies.

The manual analysis is the result of more than two full weeks of work, where we have manually analyzed dozens of sosies to investigate what kind of software diversity results from sosiefication. 
At a very coarse grain, before explaining them in details, we distinguish three kinds of sosies:
\begin{inparaenum}[(i)]
  \item \emph{revealer sosies} indicate the presence of software plasticity in the code; 
  \item \emph{fooler sosies} are named after Cohen's \cite{cohen93} counter-measures for security. 
  \item \emph{buggy sosies} are made on transplantation points that are poorly specified by the test suite, the transformation simply introduces a bug.
\end{inparaenum}

\emph{Revealer sosies} take their denomination from the fact that they reveal something in the code that is implicit otherwise. In the context of software diversification, they reveal the presence of forgiving regions. 
Once those regions are revealed,  a diversification algorithm can target them, with a high confidence that the variant will be acceptable.    

\emph{Fooler sosies} are called like this in reference to the ``garbage insertion'' transformation proposed by Cohen \cite{cohen93}. These sosies add garbage code that can fool attackers who look for specific instruction sequences. To this extent,  sosiefication  can be seen as a realization of Cohen's transformation.

\emph{Buggy sosies} are simply the degenerated and uninteresting by-products resulting from of weak test cases.  We will not provide a taxonomy of buggy sosies. 

In the following, we  discuss categories of revealer and fooler sosies. For each category, we 
present a single archetypal example from the ones synthesized for this work (table \ref{tab:sosies}). 
Each example illustrates the difference in the original that produces a sosie. Examples come with a table that provides the values for  the transplantation point features. A more complete set of examples is available online\footnote{\scriptsize \url{github.com/DIVERSIFY-project/sosie-dataset}}.

\textbf{Plastic specification.} Some program regions implement behavior which correctness is not binary. In other terms, there is no one single possible correct value, but rather several ones. We call such specification ``plastic''. 
The regions of code implementing plastic specifications are  extremely forgiving.
They provide great opportunities for sosiefication which transforms the programs in many ways  while maintaining valuable and correct-enough functionality. 

One situation that we have encountered many times relates to the production of hash keys. 
Methods that produce these keys have a very plastic specification: they must return an integer value that can be used to identify an element. The only contract  is that the function must be deterministic. 
Otherwise, there is no other constraint on the value of the hash key. 
Listing \ref{lst:coll-hash} illustrates an example of a sosie synthesized by removing a statement from a hash method (line  \ref{hash-removed}). To us, the sosie still provides a perfectly valid functionality.

\begin{figure}[ht]
\begin{lstlisting}[caption={Delete a statement in \texttt{hash} (commons.collection)},label={lst:coll-hash},numbers=left]
int hash(final Object key) {
_-    int h = key.hashCode();_ %* \label{hash-removed} *)
    h += ~(h << 9); 
    h ^=  h >>> 14;
    h +=  h << 4;
    h ^=  h >>> 10;
    return h;}
\end{lstlisting}
\tabcolsep=0.11cm
\scriptsize
\begin{tabular}{>{\small}c>{\small}c>{\small}c}
\hline
\rowcolor{lightgray} \#tc &   transfo type & node type   \\
\hline
 422 &  del & var declaration  \\
\hline
\end{tabular}
\end{figure}

\textbf{Optimization} Some code is pure optimization, which is an ideal forgiving regions for diversification. If one removes it, the output is still exactly the same, only non-functional properties such as performance are impacted. 
Listing \ref{lst:range-tostring} shows an example of sosie that removes an optimization: at the end of the \texttt{if-block} (line \ref{tostring-removed}), the original program stores the value of \texttt{buf} in  \texttt{toString}, which allows to bypass the  computation of buf next time \texttt{toString()} is called; the sosie removes this part of the code, producing a potential performance degradation if the method is called intensively.

\begin{figure}[ht]
\begin{lstlisting}[caption={Delete a statement in \texttt{toString} (commons.lang)},label={lst:range-tostring},numbers=left]
String toString() {
  String result = toString;
  if (result == null) {
    final StringBuilder buf = new StringBuilder(32);
    %*{\textbf{\color{gray}...compute buf}}*)
    result = buf.toString();
_-    toString = result;_ %* \label{tostring-removed} *)  
  }
  return result;}
\end{lstlisting}
\tabcolsep=0.11cm
\scriptsize
\begin{tabular}{>{\small}c>{\small}c>{\small}c}
\hline
\rowcolor{lightgray} \#tc &   transfo type & node type   \\
\hline
2&   del &stmt list  \\
\hline
\end{tabular}
\end{figure}

\textbf{Code redundancy.} It sometimes happens that the very same computation is performed several times in the same program. For instance, two subsequent calls to \texttt{list.remove(o)}, even separated by  other instructions are equivalent (as long as \texttt{list} and \texttt{o} do not change between).
Sosiefication naturally exploits this  computation redundancy through the removal or replacement of these redundant statements. 
Replacement with side-effect free also produces valid sosies.

Listing \ref{lst:center} displays an example of such a sosie  (removing if-block at line \ref{center-removed}). 
The statement \texttt{if (isEmpty(padStr)) { 
  padStr = SPACE;
} }  assigns a value to  \texttt{padStr}, then this variable is passed to methods \texttt{leftPad} and \texttt{rightPad}. Yet, each of these two methods include the exact same statement, which will eventually assign a value to \texttt{padStr}. So, the statement is redundant and can be removed from the original program, yielding a valid fooler sosie.
Compared to sosies that remove some optimization, those sosies might be more performant than the original program. 

\begin{figure}[ht]
\begin{lstlisting}[caption={Delete in \texttt{center} (commons.lang)},label={lst:center},numbers=left]
String center(String str, final int size, String padStr) {
  if (str == null || size <= 0) {return str;}
_-  if (isEmpty(padStr)) {padStr = SPACE;}_ %* \label{center-removed} *)
  %*{\color{gray} ... }*)
  str = leftPad(str, strLen + pads / 2, padStr);
  str = rightPad(str, size, padStr);
  return str;}
\end{lstlisting}
\tabcolsep=0.11cm
\scriptsize
\begin{tabular}{>{\small}c>{\small}c>{\small}c}
\hline
\rowcolor{lightgray} \#tc &   transfo type & node type   \\
\hline
1&  del & if \\
\hline
\end{tabular}
\end{figure}

\textbf{Implementation redundancy.} It often happens that programs embed several different functions that provide the same service, in different ways. For example, there can exist several versions of the same method with different sets of parameters, which can be used interchangeably by providing good parameter values. It is also possible to use libraries that provide this diversity of similar methods (as demonstrated by Carzaniga and colleagues \cite{Carzaniga14}). 
Listing \ref{lst:utils-get} illustrates the exploitation of such implementation redundancy inside the program (replace at line \ref{get-replace}), i.e., \texttt{((Object[]) object)[i]} has the same behavior as \texttt{Array.get(object, i)}, with completely different implementations.

\begin{figure}[ht]
\begin{lstlisting}[caption={Replace in \texttt{get} (commons.collection)},label={lst:utils-get},numbers=left]
Object get(final Object object, final int index) {
  %*{\textbf{\color{gray}...}}*)
  else if (object instanceof Object[]) {
_-    return ((Object[]) object)[i];_ %* \label{get-replace} *)
(*+    try {
+        return Array.get(object, i);
+    } catch (final IllegalArgumentException ex) {
+        throw new IllegalArgumentException("Unsupported 
+          object type: " + object.getClass().getName());
+    }*)
    } 
  %*{\textbf{\color{gray}...}}*)
}
\end{lstlisting}
\tabcolsep=0.11cm
\scriptsize
\begin{tabular}{>{\small}c>{\small}c>{\small}c}
\hline
\rowcolor{lightgray} \#tc &   transfo type & node type   \\
\hline
1 &  rep & return \\
\hline
\end{tabular}
\end{figure}

\textbf{Optional functionality.}  
In software, not all parts of equal importance. Some parts represent the core functionality, other parts are about options and are not essential to the computation. Those optional parts are either not specified or the specification is of less importance. 
These are areas that can be safely removed or replaced while still producing useful variants. 
Listing \ref{lst:canonicalize} is an example of sosie that exploits such optional functionality.
The sosie completely removes the body of the method, which is supposed to transform the type passed as parameter into an equivalent version that is serializable, and instead it returns the parameter. 
The sosie is covered by 624 different test cases, it is executed 6000 times and all executions complete successfully and all assertions in the test cases are satisfied. 
This is an example of an advanced feature implemented in the core part of GSon that is not necessary to make the library run correctly.

\begin{figure}[ht]
\begin{lstlisting}[caption={Replace in \texttt{canonicalize} (GSon)},label={lst:canonicalize},numbers=left]
public static Type canonicalize(Type type) {
_-  if (type instanceof Class) { 
-    Class<?> c = (Class<?>) type;
-    return c.isArray() ? new
-    GenericArrayTypeImpl(canonicalize(c.getComponentType())) : c;
-    }
-  else
-    if (type instanceof ParameterizedType) {
-      ParameterizedType p = (ParameterizedType) type;
-      return new ParameterizedTypeImpl(p.getOwnerType(),
-          p.getRawType(), p.getActualTypeArguments());
-    } 
-    else 
-      if (type instanceof GenericArrayType) {
-        GenericArrayType g = (GenericArrayType) type;
-        return new -GenericArrayTypeImpl(g.getGenericComponentType());
-     } 
-      else 
-      if (type instanceof WildcardType) {
-        WildcardType w = (WildcardType) type;
-        return new WildcardTypeImpl(w.getUpperBounds(),   
-            w.getLowerBounds());
-      } 
-      else {
-        return type;
-      } _
(*+  return type;*)
}
\end{lstlisting}
\tabcolsep=0.11cm
\scriptsize
\begin{tabular}{>{\small}c>{\small}c>{\small}c}
\hline
\rowcolor{lightgray} \#tc &   transfo type & node type   \\
\hline
623 &  rep & if   \\
\hline
\end{tabular}
\end{figure}

\begin{figure}[ht]
\begin{lstlisting}[caption={Add in \texttt{ensureCapacity} (commons.collection)},label={lst:ensureCapacity},numbers=left]
void ensureCapacity(final int newCapacity) {
  final int oldCapacity = data.length;
  if (newCapacity <= oldCapacity) {
    return;
  }
  if (size == 0) {
    threshold = calculateThreshold(newCapacity, loadFactor);
    data = new HashEntry[newCapacity];
  } else {
    %*{\color{gray}...}*)
    }
(*+  ensureCapacity(threshold)*)}%* \label{ensure-add} *)
\end{lstlisting}
\tabcolsep=0.11cm
\scriptsize
\begin{tabular}{>{\small}c>{\small}c>{\small}c}
\hline
\rowcolor{lightgray} \#tc &   transfo type & node type   \\
\hline
 8&   add & invocation \\
\hline
\end{tabular}
\end{figure}

\textbf{Fooler sosies.}

We have realized that a number of ``add'' and ``replace'' transformations result in sosies which have more code than the original and where the additional code is harmless for the overall execution. 
These sosies act exactly as Cohen's ``garbage insertion'' strategy to fool malicious attackers, hence we call them fooler sosies.

We found multiple kinds of fooler sosies: some add branches in the code or redundant method calls or redundant sequences of method calls. Some others reduce the legitimate input space through additional checks on input parameters.
Listing \ref{lst:ensureCapacity} is an example of a fooler sosie, which adds a recursive call to \texttt{ensureCapacity()} (line \ref{ensure-add}). 
This could turn the method into an infinite recursion, except that in the additional recursive call, the value of the parameter is  such that the condition of the first if-statement  always holds true and  the method execution immediately stops.
The additional call adds a harmless method call in the execution flow.

\textbf{Discussion} 
Let us now consider again the transplantation point features given for each sosie. 
Most sosies identified as buggy with we manual analysis are done on transplantation points covered by a single test case. 
In other words,  the risk of synthesizing bad sosies increases when the number of test cases is low.

More interestingly, we realized that valid revealer and fooler sosies can be found both on points intensively tested and on weakly tested points.
This makes us conclude that if a region is intrinsically plastic (has a plastic specification or is optional), the number of test cases barely matters, the only fact that the specification and the corresponding code region is plastic explains the fact that we can easily synthetize sosies. 
This  confirms a trend we observed in RQ1: no matter how much a region is tested, we can synthesize sosies because of some intrinsic forms of plasticity.

\begin{framed}
Answer to RQ3: We have provided a first classification or software sosies, founded on the concepts of revealer, fooler and buggy sosies. The ``revealers'' indicate forgiving regions \cite{rinard2012}. The ``foolers'' are useful in a protection setting \cite{cohen93}. The buggy sosies are due to weak test cases.
Our manual analysis shows the variety of roles that code plays in a program. It uncovers the multitude of opportunities that exist for sosie synthesis and diversification in real-world programs.
\end{framed}

\subsection{Threats to Validity}
\label{sec:threats}

We performed a large scale experiment in a relatively unexplored domain: software diversification at the application code level.
We now present the threats to the validity. 

Our findings might not generalize to all types of applications. 
We selected frameworks and libraries because of their popularity, their longevity and the very high quality of their test suites. 
Yet, our observations about the large variations among statements, with respect to test coverage, and about code plasticity can be different when analyzing programs in other domains. 

Our large scale experiments rely on a complex tool chain, which integrates code transformation, instrumentation, trace analysis and statistical analysis. 
We also rely on the Grid5000 grid infrastructure to run  millions of transformations.
We did extensive testing of our code transformation infrastructure, built on top of the Spoon framework that has been developed, tested and maintained for over more than 10 years.
However, as for any large scale experimental infrastructure, there are surely bugs in this software. We hope that they  only change marginal quantitative things, and not the qualitative essence of our findings. Our infrastructure is publicly available on Github \footnote{\scriptsize \url{github.com/DIVERSIFY-project/sosie-dataset}}.

\section{Related Work}
\label{sec:rw}

As mentioned on several occasions in this paper, our work is related to the multiple investigations of Martin Rinard and his group about software tradeoffs between correctness and other properties such as security or performance. 
Rinard has defined the general concept of ``acceptability envelop'', and explored its application in different domains.
For example, they injected off-by-one errors on loop termination conditions in order to characterize the behavior of two programs under errors \cite{rinard2005}, they also experimented with runtime loop perforation to explore the same envelop \cite{sidiroglou2011}. 
In all these cases, the authors use a set of test scenarios to assess the acceptability of the changes. 
Our work contributes to this body of knowledge about the nature of the acceptability envelop by investigating new kinds of transformations  as well as a new analysis method to locate code regions that can tolerate changes. The set of \emph{revealer} and \emph{fooler} sosies for a given program can be considered as forming the body within the ``acceptability envelop'' of the program \cite{rinard2005}. 

Mutational robustness \cite{Schulte13} is the ability of software to resist to mutations. 
The essential difference between both works lies in the definition of program transformations: 
Schulte et al. use only random operations, while we use a heuristics based on types and variable renaming.
Also, Schulte et al. say that software is robust to mutations, we say that we can synthesize diversity and that this indicates the presence of true plasticity in the code. 

The recent advances in software transplantation by Sidiroglou and colleagues \cite{sidiroglou15} and Barr and colleagues \cite{Barr15} is related to sosiefication. 
Both work transfer code from a donor program into recepient applications. 
Sidiroglou performs transplantation for bug fixing purposes and Barr does it to reuse functionality from one program to another.
Sosiefication, especially the fooler sosies, can be seen as a form of internal micro transplantation.

The work of Langdon and Harman \cite{langdon2013} defines an iterative process of code transformations and testing in order to speed-up program execution. Schulte and colleagues use a similar process to reduce energy consumption of embedded programs \cite{schulte2014}. 
Works in the area of genetic improvements of programs is related to ours since they also rely on code transformations and test suites in order to automatically produce different versions of a program.
Our analysis of statement execution signatures could also improve such approaches.


Our investigations of software plasticity at the edge of correctness tradeoffs directly relate to seminal works that advocate for novel ways of building software that is more approximate and evolvable, but also less brittle. 
In particular, our work is very much inspired by the work of Richard Gabriel  \cite{gabriel2006}, Gerald Sussman \cite{sussman07} and Mary Shaw \cite{shaw02}. They all warn against the desire of building perfectly correct system, which can only be correct in very specific conditions and are consequently very brittle outside these conditions. They advocate for new approaches that would support the construction of software systems that have the ability to evolve and adapt, in exchange of certain tradeoffs with respect to correctness. 
We foresee our investigations about automatic diversification of application source code as a contribution towards the design of such new approaches.

\section{Conclusion}
\label{sec:conclusion} 

In this paper, we have presented an exploration in the area of software diversification.
We have analyzed a specific diversification technique -- sosiefication  -- in the light of the interactions between a test suite and the program under test. 
This investigation combined automated analysis with the  manual exploration of a large sample of sosies. 
This enabled us to contribute to the body of knowledge on automatic software diversity as follows.
First, we have shown the correlation between statement execution signatures and sosiefication, and we demonstrated that sosiefication rate never goes down to zero, indicating a certain degree of intrisic plasticity in any program;
Second, we have provided novel pieces of evidence about the presence and the nature of forgiving regions in software. 
Third, we demonstrated the effectiveness of code addition and deletion, to synthesize sosies that can contribute to previous work on OS protection by Cohen \cite{cohen93} and failure oblivious computing by Rinard \cite{rinard2012}.

As future work, we wish to exploit these findings in order to automate the synthesis of variants that establish tradeoffs between functional correctness and other qualities such as performance. 
We believe that software developers must constantly take into account a wide variety of concerns into the code that goes into production and, to this extent, they must constantly take multi-criteria decisions. 
Eventually they deliver a product that is a single point on the Paretto of all possible solutions that can satisfy the same requirements. 
We want to exploit sosiefication and other diversification techniques as a way to automatically explore the neighbourhood on this Paretto front.

\bibliography{references}
\bibliographystyle{plain}

\end{document}